\documentclass[%
 reprint,
superscriptaddress,
 amsmath,amssymb,
 aps,
 floatfix,
pra,
]{revtex4-2}

\usepackage{graphicx}% Include figure files
\usepackage{dcolumn}% Align table columns on decimal point
\usepackage{bm}% bold math
\usepackage{siunitx}
\usepackage{amsmath}
\usepackage[]{hyperref}
\usepackage{xcolor}
\usepackage{braket}
\usepackage{graphicx}% Include figure files
\usepackage{dcolumn}% Align table columns on decimal point
\usepackage{bm}% bold math
\usepackage[mathlines]{lineno}% Enable numbering of text and display math
\usepackage{ulem}
\usepackage[T1]{fontenc}
\usepackage[utf8]{inputenc}
\begin{document}
\preprint{APS/123-QED}

% Use the \preprint command to place your local institutional report
% number in the upper righthand corner of the title page in preprint mode.
% Multiple \preprint commands are allowed.
% Use the 'preprintnumbers' class option to override journal defaults
% to display numbers if necessary
%\preprint{}

%Title of paper
\title{Altermagnetic nanotextures revealed in bulk MnTe}

\author{Rikako Yamamoto}
\email{rikako.yamamoto@cpfs.mpg.de}
\affiliation{Max Planck Institute for Chemical Physics of Solids, 01187 Dresden, Germany}
\affiliation{International Institute for Sustainability with Knotted Chiral Meta Matter (WPI-SKCM$^{2}$), Hiroshima University, Hiroshima 739-8526, Japan}

\author{Luke Alexander Turnbull}
\affiliation{Max Planck Institute for Chemical Physics of Solids, 01187 Dresden, Germany}
\affiliation{International Institute for Sustainability with Knotted Chiral Meta Matter (WPI-SKCM$^{2}$), Hiroshima University, Hiroshima 739-8526, Japan}
\author{Marcus Schmidt}
\affiliation{Max Planck Institute for Chemical Physics of Solids, 01187 Dresden, Germany}
\author{Jos\'{e} Claudio Corsaletti Filho}
\affiliation{Max Planck Institute for Chemical Physics of Solids, 01187 Dresden, Germany}
\author{Hayden Jeffrey Binger}
\affiliation{Max Planck Institute for Chemical Physics of Solids, 01187 Dresden, Germany}
\author{Marisel Di Pietro Mart\'inez}
\affiliation{Max Planck Institute for Chemical Physics of Solids, 01187 Dresden, Germany}
\affiliation{International Institute for Sustainability with Knotted Chiral Meta Matter (WPI-SKCM$^{2}$), Hiroshima University, Hiroshima 739-8526, Japan}
\author{Markus Weigand}
\affiliation{Institute for Nanospectroscopy, Helmholtz-Zentrum Berlin, 12489 Berlin, Germany}
\author{Simone Finizio}
\affiliation{Institute for Nanospectroscopy, Helmholtz-Zentrum Berlin, 12489 Berlin, Germany}
\affiliation{Swiss Light Source, Paul Scherrer Institute, 5232 Villigen PSI, Switzerland}
\author{Yurii Prots}
\affiliation{Max Planck Institute for Chemical Physics of Solids, 01187 Dresden, Germany}
\author{George Matthew Ferguson}
\affiliation{Max Planck Institute for Chemical Physics of Solids, 01187 Dresden, Germany}
\author{Uri Vool}
\affiliation{Max Planck Institute for Chemical Physics of Solids, 01187 Dresden, Germany}
\author{Sebastian Wintz}
\affiliation{Institute for Nanospectroscopy, Helmholtz-Zentrum Berlin, 12489 Berlin, Germany}
\author{Claire Donnelly}
\email{claire.donnelly@cpfs.mpg.de}
\affiliation{Max Planck Institute for Chemical Physics of Solids, 01187 Dresden, Germany}
\affiliation{International Institute for Sustainability with Knotted Chiral Meta Matter (WPI-SKCM$^{2}$), Hiroshima University, Hiroshima 739-8526, Japan}

\date{\today}
\begin{abstract}
Altermagnetism represents a magnetic phase in which the combination of compensated antiferromagnetic order with an anisotropic crystal field leads to time reversal symmetry breaking. The resulting combination of properties typically associated with ferromagnets, but with net-zero magnetisation has generated significant interest - both for fundamental research, and technological applications.
With many candidate altermagnetic materials, MnTe has emerged as one of the most promising systems, with growing experimental evidence for altermagnetic phenomena. {So far,} the majority of measurements {have been} performed on thin-films, or {have involved} surface measurements. {However,} the question of altermagnetic order in the bulk system - in the absence of substrate or surface effects - remains. 
Here we show evidence for bulk altermagnetism in single crystal MnTe, through spectroscopic X-ray microscopy. By performing nanoscale X-ray magnetic circular dichroic (XMCD) imaging in transmission on a $200\,\rm{nm}$-thick lamella, we observe domains and magnetic textures with a spectroscopic signature characteristic of altermagnetic order, thereby confirming the intrinsic nature of altermagnetism in MnTe. Quantitative analysis of the XMCD signal reveals an excellent agreement with predicted signals, establishing that the altermagnetic order exists throughout the thickness of the lamella and confirming the {intrinsic, bulk} nature of the state. With these results, we demonstrate that transmission XMCD spectroscopic imaging is a robust, quantitative technique to probe altermagnetic order, providing a means to probe individual altermagnetic domains within complex configurations.
{This ability to investigate, and characterise altermagnetic order in bulk crystals represents an important tool for the exploration of altermagnetism across a wide range of candidate materials, of key importance for the development of future technologies.}
\end{abstract}

\maketitle
\section{\label{sec:Introduction}Introduction}
{Magnetism has played an important role in society, from  ferromagnets that have led to the generation of electricity and more recently facilitated the information age, to the more recently discovered antiferromagnets {that offer a route to robust, THz technologies}. When it comes to spintronic applications, identifying the optimal type of magnetic order is important \cite{Bai2024}.
On the one hand, ferromagnets exhibit strong spintronic effects {such as the anomalous Hall effect (AHE) and giant magnetoresistance}, allowing their magnetic configuration to be efficiently read and written \cite{Kim2024}. 
However, their net magnetization makes them susceptible to external magnetic fields, and their dynamic response, which determines information transfer speeds, is limited to the GHz range \cite{Giustino_2020}.
On the other hand, antiferromagnets have zero net magnetization, making them highly robust and enabling technologically relevant THz dynamics \cite{Cheng2016, Hongsong2023}. 
However, despite first demonstrations of all-electrical reading and writing of antiferromagnetic order \cite{jungwirth2016AFspintronics}, their spintronic signals are generally weak, limiting their potential applications.}

\begin{figure*}
\includegraphics{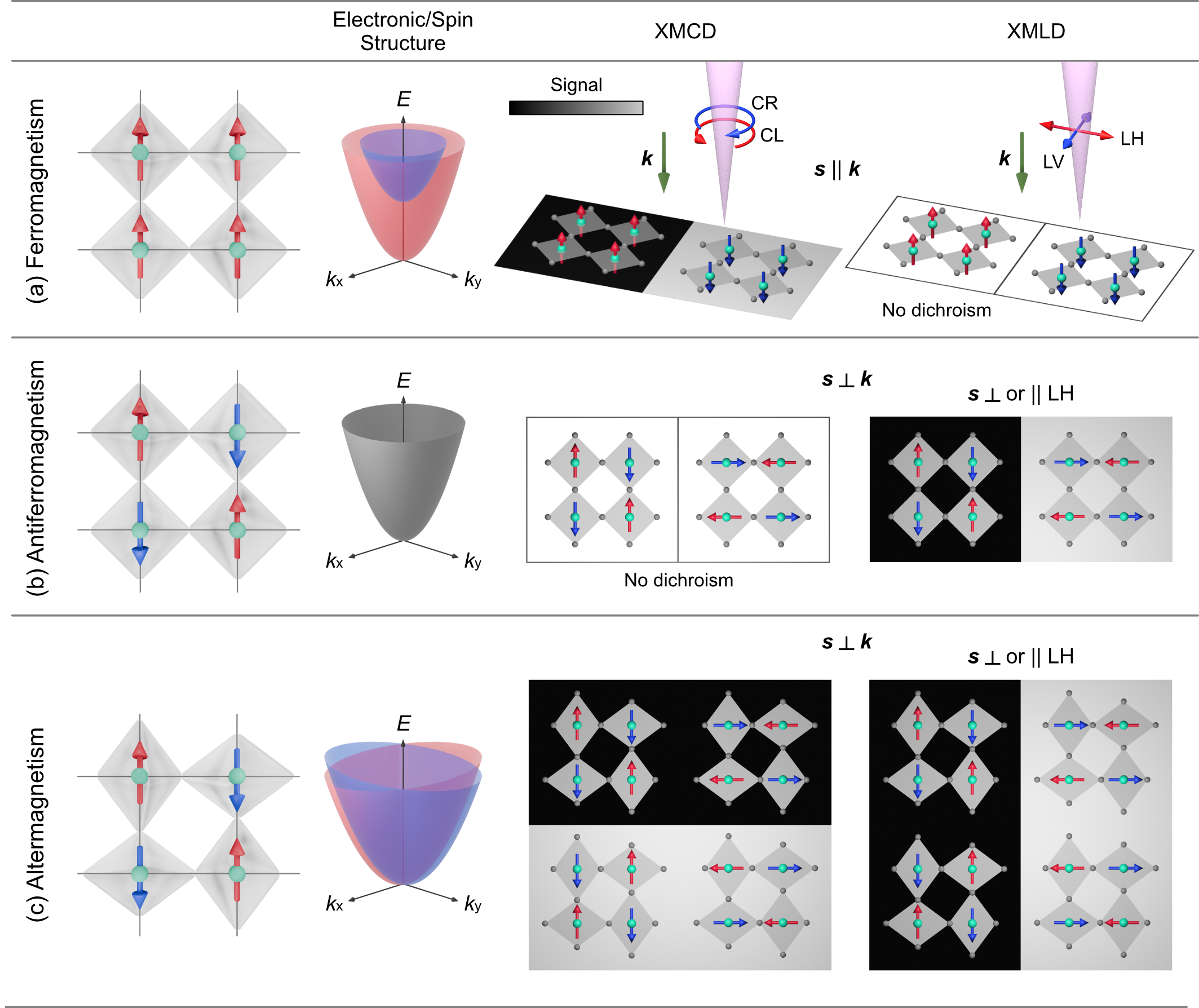}%
\caption{\label{fig:fig1} { Mapping magnetic domains {in ferromagnets, antiferromagnets and altermagnets} using polarized X-rays.
The polarisation-dependent transmission can be used to identify and map ferromagnetic, antiferromagnetic, and altermagnetic domains.
The spin configuration and local crystal environments  of ordered magnetic ions (left), the electronic and spin structure in momentum space (centre) and the expected XMCD and XMLD contrast for different domain configurations (right) is shown for ferromagnetism (a), antiferromagnetism (b) and  $d$-wave altermagnetism (c). Throughout, red and blue indicate opposite spin orientations, while the black, and white, and grey planes in the XMCD/XMLD schematics indicate positive, negative and zero XMCD or XMLD intensity, respectively.
(a) In ferromagnets, neighbouring spins have parallel alignment, associated with a well-defined splitting in momentum space. For the spins orientated out of the plane, collinear to the direction of the X-rays ($\bm{s} \parallel \bm{k}$), there is a non-zero X-ray magnetic circular dichroism (XMCD), but no X-ray magnetic linear dichroism (XMLD). 
(b) For collinear antiferromagnets with in-plane spin configuration ($\bm{s} \perp \bm{k}$), neighbouring spins have anti-parallel alignments, with no associated spin splitting in momentum space. There is no XMCD signal associated with this state, however XMLD distinguishes between perpendicular spin configurations. 
(c) In d-wave altermagnets shown here, at least four types of magnetic domains form due to the combination of antiferromagnetic spin configurations, and the inequivalent local environments. (Note that for the case of MnTe, there are six different types of magnetic domains with respect to the easy axes.) For the case of X-rays propagating perpendicular to the plane in which the spins lie, there is both non-zero XMCD - that probes the relative orientation of the spins with respect to the local environment, and linear dichroism, which probes the relative orientation of the spins with respect to the linear polarisation of the light. A comprehensive overview of XMLD and XMCD for different spin orientations is given in the Appendix {Figure \ref{fig:fig6}. }}
}
\end{figure*}

Altermagnets provide a possible solution to this problem \cite{Smejkal2022_PRX_first_obs,vsmejkal2022altermagnets}, as they combine features of ferro- and antiferromagnets {(see Figure \ref{fig:fig1})} that are of importance for potential spintronic applications. 
In particular, an altermagnet has a collinear compensated spin configuration, which is associated, as in {broader} antiferromagnets, with net zero magnetisation, and ultra-fast dynamics on the sub-picosecond timescale \cite{Gray2024}. 
However, compared to antiferromagnets, there is an additional symmetry breaking associated with the local environment of each sublattice, as shown in Figure \ref{fig:fig1} (c). This gives rise to unconventional spin splitting of the bands in momentum space, which in turn results in measurable spintronic signals such as AHE and tunneling magnetoresistance effects, thus promising to combine the robust, ultrafast nature of antiferromagnets with ferromagnetic-like phenomena, representing an opportunity for future technological applications \cite{dal_din_antiferromagnetic_2024}.

This concept of altermagnetism as a third form of ferroic magnetic order was first defined by considering the spin and crystal symmetries separately \cite{Smejkal2022_PRX_first_obs}, which allowed for a complete classification of all the possible spin arrangements on crystals, and the identification of altermagnetic phases with $d$-, $g$-, and $i$-wave spin order in momentum space. This discovery was motivated by previous observations of anomalies such as Hall effects in collinear antiferromagnetic order \cite{Smejkal2020_SciAdv,Samanta2020_JAP,Naka2020_PRB,Feng2022}, spin-current generation \cite{Naka2019_NatComms,Gonzalez2021_PRL} and unconventional spin-splittings \cite{Ahn2019,Hayami2019_JPSJ,Yuan2020_PRB, Hayami2020_PRB, Yuan2021_PRM}.
Since the first prediction of altermagnetism \cite{Smejkal2022_PRX_first_obs}, these symmetry arguments have been used to predict a wide variety of known – and unknown – materials to be altermagnetic. 
This has led to renewed interest in materials that in some cases have long been studied, specifically now looking for evidence of altermagnetism. 
Indeed, first experimental indications of altermagnetism were given by the presence of an AHE in a variety of compensated magnets \cite{Smejkal2020_SciAdv,Gonzales_AHE_MnTe,Feng2022,Reichlova2024_NatComms,Kluczyk2024}, in which such an effect would not traditionally be expected \cite{Smejkal2022_NatMat}. 
Direct evidence of time reversal symmetry breaking was provided by the band splitting in momentum space measured by angle-resolved photoemission spectroscopy (ARPES) \cite{Krempasky2024}.  {Subsequent measurements of X-ray magnetic circular dichroism  (XMCD) in altermagnets \cite{hariki_x-ray_2024}, again an effect that would not traditionally be expected for antiferromagnetic materials, allowed for the direct imaging of altermagnetic nanoscale order – both domains, domain walls and topological textures \cite{amin_altermagnets_2024}. Indeed, the XMCD itself provides evidence of time-reversal symmetry breaking, not only providing access to nanoscale configurations, but allowing one to probe the altermagnetic nature of the material.}

{With the growing number of theoretical and experimental works, the next challenge lies in identifying and confirming candidate altermagnetic materials.} Initially, studies focused on the candidate $\mathrm{RuO_2}$ among several proposed candidates \cite{smejkal_AHE_2020}, a collinear antiferromagnet with the rutile structure. Despite first measurements in thin films that suggested antiferromagnetic ordering \cite{Berlijn2017, Zhu2019}, the presence of AHE \cite{Feng2022} and spin splitting \cite{Fedchenko2024, Liu2024} characteristic of altermagnets, it has remained challenging to conclusively determine the nature of the material in the bulk \cite{Hiraishi2024, Kesler2024, JLiu2024}. 
{Recently, MnTe has  emerged as the next promising candidate. First measurements on thin film systems grown by molecular beam epitaxy (MBE) \cite{amin_altermagnets_2024, Gray2024, Krempasky2024, Lee2024_ARPES} {revealed} the successful measurement of spin-splitting in momentum space \cite{Krempasky2024,Lee2024_ARPES}, as well as the existence of circular dichroism \cite{hariki_x-ray_2024,amin_altermagnets_2024}. While the {vast majority of transport measurements} of the AHE have been mainly focused on thin film systems \cite{Gonzalez2023_PRB}, recent measurements of bulk MnTe have also reported the AHE, albeit with certain discrepancies with respect to thin film systems \cite{Kluczyk2024}.
{As well as these discrepancies between the bulk and the thin film measurements, i}ndications of the influence of substrate-induced strain have been observed both in transport measurements  \cite{Bey2024}, as well as in nanoscale XMCD imaging, where lithographic patterning appears to result in local strain relaxation for the manipulation of the order \cite{amin_altermagnets_2024}.  
In this context, disentangling the role of the substrate or surface in thin films or surface-sensitive measurements, and the resulting local strain and symmetry breaking, is highly important to clarifying the intrinsic altermagnetic nature of MnTe \cite{Chakraborty2024, Chen2024,Osumi2024}.

\begin{figure}
\includegraphics{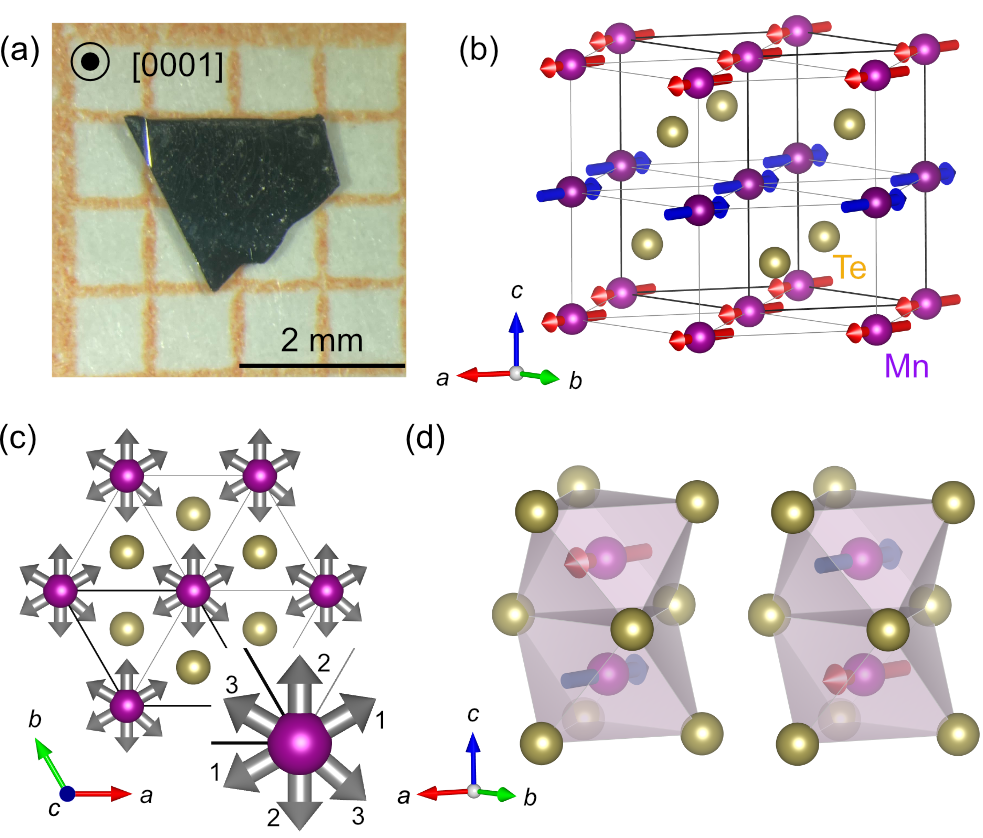}%
\caption{\label{fig:fig2} 
{Crystal structure and magnetic order of MnTe. (a) Bulk MnTe crystal grown with chemical vapour transport. (b) Crystallographic and magnetic structure of MnTe (c) Three easy antiferromagnetic spin axes of MnTe.  (d) Taking the three dimensional crystal structure surrounding the Mn atoms into account, one can distinguish altermagnetic domains with opposite sublattice orientation that are related by time reversal symmetry. }
}
\end{figure}

\begin{figure}
\includegraphics{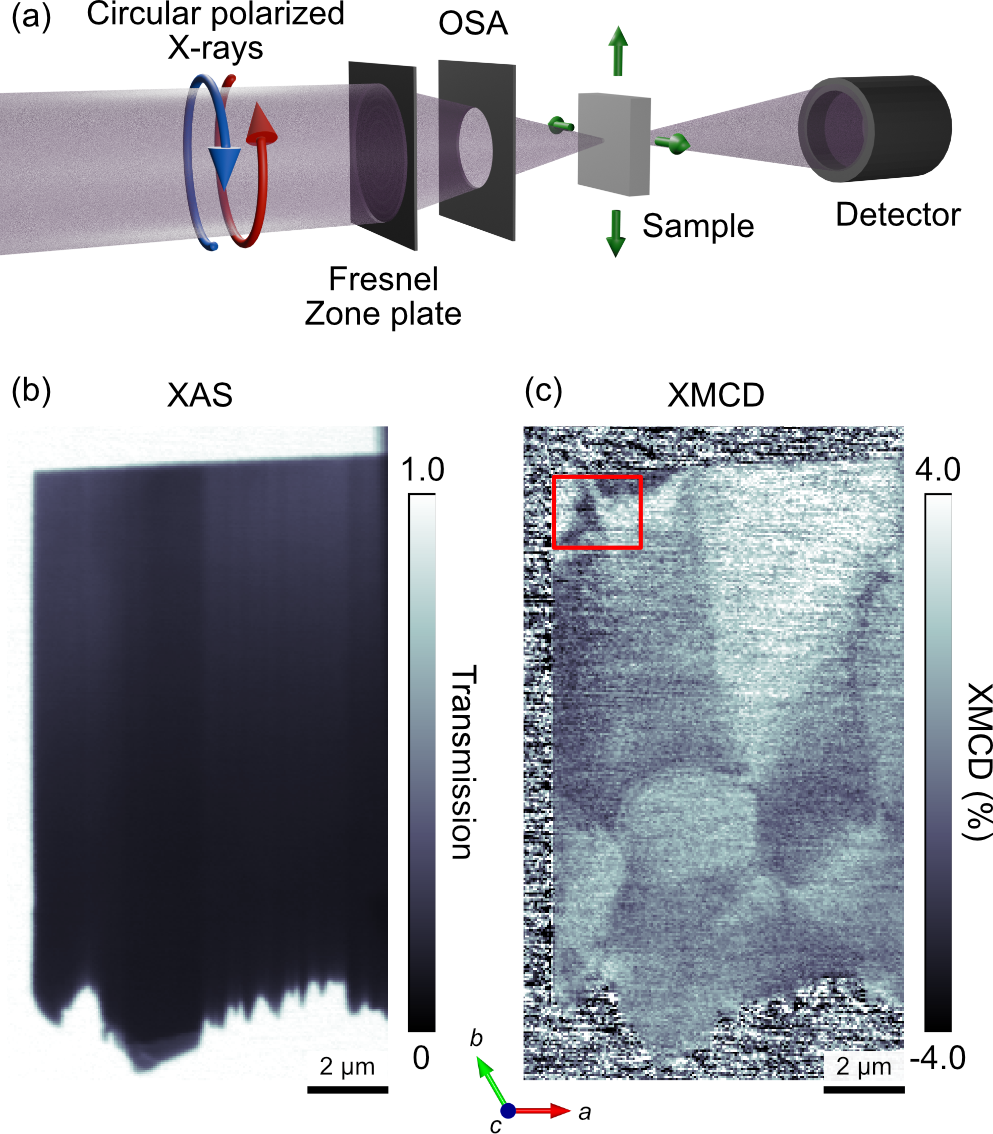}%
\caption{\label{fig:fig3}
{X-ray magnetic circular dichroic scanning transmission X-ray microscopy of altermagnetic domains in MnTe.  (a) Schematic of Scanning Transmission X-ray microscopy setup, where circularly polarised X-rays are focused onto a sample that is scanned to acquire a transmission projection with nanoscale resolution. (b) A single polarisation projection of the lamella, and (c) an XMCD projection of the lamella, showing different regions with dark and bright contrast indicating the presence of domains and magnetic features.} }
\end{figure}

Here we obtain experimental evidence of bulk altermagnetism in a lamella extracted from a single crystal of MnTe with nanoscale X-ray spectroscopic imaging.
We observe magnetic domains and textures with X-ray magnetic circular dichroism {that exhibits the characteristic three-fold symmetry expected for MnTe. }We measure an associated spectroscopic XMCD signal that changes sign multiple times across the Mn $L_{2,3}$ edges.
This oscillating signal is characteristic of altermagnetic MnTe, and exhibits remarkable quantitative agreement with theoretical calculations \cite{hariki_x-ray_2024}.  
By measuring a free-standing lamella in transmission, we integrate through the material, {providing evidence that altermagnetic order exists in the absence of} surface effects and substrate-induced strain. 
This, combined with the quantitative agreement of the measured dichroism with theoretical calculations,  indicates that the altermagnetic order exists within the bulk of the material, thus providing evidence for the intrinisic nature of altermagnetism in MnTe. 
Such transmission XMCD measurements represent an important tool to understand altermagnetic order in bulk samples and open up the \textit{in situ} study of altermagnets under the application of external stimuli such as magnetic fields and currents, both for the probing of fundamental characteristics, and for the development of future technologies. 

\section{XMCD in altermagnets}

To probe the altermagnetic nature of MnTe in the bulk, we turn to X-ray magnetic circular dichroism. An overview of magnetic dichroism for different types of magnetic order -- ferro-, antiferro- and alter- -- is given in Figure \ref{fig:fig1}, with a comprehensive overview given in the Appendix. Commonly associated with ferromagnets, XMCD represents the difference in the scattering factor {and absorption cross section} of X-rays for circular left and right handed light, in the vicinity of an absorption edge. 
For collinear antiferromagnets, which have zero net magnetisation, there is typically no circular dichroism. 
Instead, linear magnetic dichroism, where the scattering factor depends on the relative orientation of the polarisation of the light and the Néel vector, can be used to probe the local order, as shown in Figure \ref{fig:fig1}(b). 
Remarkably, altermagnets not only exhibit the expected linear dichroism due to the orientation of the Néel vector, but also exhibit XMCD (see Figure \ref{fig:fig1}(c)). This circular dichroism is not due to a net magnetic moment in the system, but is of a different origin from the ferromagnetic case. 

The origin of XMCD for altermagnetic materials is {generally} associated with the time reversal symmetry breaking, and has been understood in the context of the Hall vector which the XMCD effectively probes \cite{hariki_x-ray_2024, amin_altermagnets_2024}.
We can also understand the origin of this XMCD by considering the nature of XMCD in more detail. 
While XMCD is commonly associated with spin, in reality XMCD actually probes an effective spin that comprises multiple terms \cite{Carra1993}. 
First, XMCD probes the  spin and orbital angular momenta parallel to the X-ray direction, which can be separated quantitatively using the XMCD sum rules \cite{Thole1992, OBrien1994}. 
However, when measuring XMCD, there is an additional contribution to the effective spin probed, known as the ``anisotropic magnetic dipole'', commonly referred to as $\braket{T_z}$. 
This originates from the anisotropic distribution of spin in the probed band \cite{Piamonteze2009}.
In typical XMCD experiments, this is often neglected, and indeed, is mostly considered when accurately calculating the spin and orbital moments of a material with sum rules \cite{Collins_1995, Sipr2018}. 
For a simple collinear antiferromagnet, the two antiparallel-aligned magnetic sublattices are equivalent, and  the pure spin, as well as the effective magnetic dipole,  therefore cancel to zero. However, in an altermagnet, where spins of opposite direction have different local environments, the additional $\braket{T_z}$ term cannot necessarily be neglected \cite{Sasabe23_TzRuO2}, and can lead to non-zero XMCD, as shown schematically for a simple $d$-wave altermagnet in Figure \ref{fig:fig1}.  
This anisotropic magnetic dipole - or $\braket{T_z}$ term - has been shown to be necessary for emergent ferromagnetic-like behaviour in antiferromagnets, such as the anomalous Hall effect \cite{Hayami21_AMD_AHE}.
For the case of MnTe it has been shown that these different origins of circular dichroism – that associated with the net magnetisation, and the dichroism associated with the altermagnetic order – can be distinguished by their spectroscopic nature \cite{hariki_x-ray_2024}. 
As a result, spectroscopic measurements of XMCD offers a route to probe the altermagnetic order of materials.
\begin{figure*}
\includegraphics{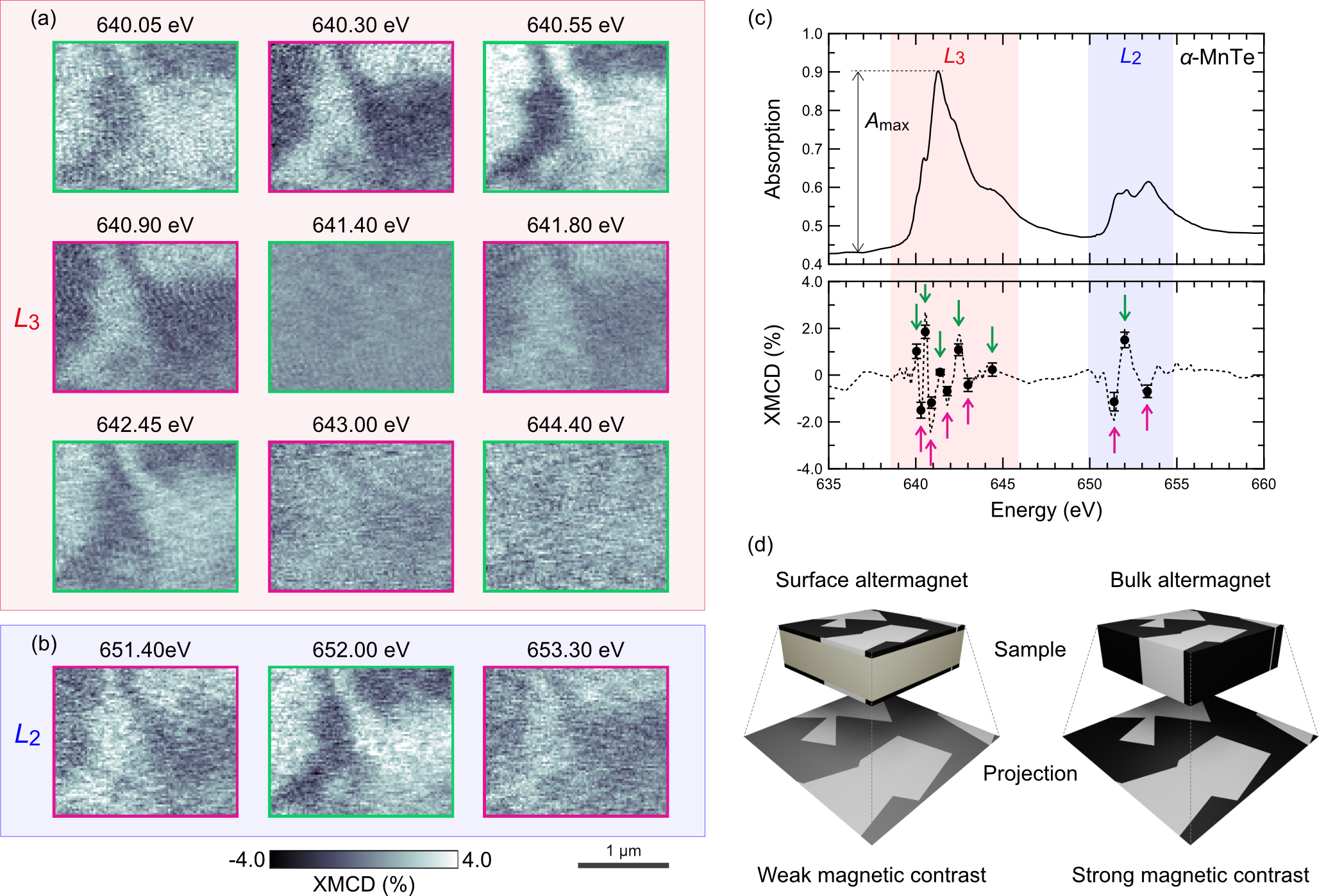}%
\caption{\label{fig:fig4} XMCD spectromicroscopy across the Mn $L_{\mathrm{2,3}}$ edges. (a) A zoomed region of the lamella (corresponding to the box in Figure \ref{fig:fig3}(c)) is shown for 9 energies across the $L_{\mathrm{3}}$ edge, across which the contrast can be seen to switch eight times. Note that lower contrast is observed for energies on the absorption edge ($641.4\,\rm{eV}$ and $641.8\,\rm{eV}$) that is likely due to high absorption of the sample. (b) The same region is shown for three energies across the $L_{\mathrm{2}}$ edge, where the contrast can be seen to switch twice.  (c) The X-ray absorption spectrum (solid line) and the XMCD spectrum (dashed line) acquired on a single domain within the lamella are plotted, along with the XMCD contrast extracted from the images shown in (a,b) (black dots, highlighted by green and purple arrows). (d) Schematic of the expected contrast of (left) surface and (right) bulk altermagnetic order in transmission. The quantitative agreement of the XMCD signal with DFT calculations presented in \cite{hariki_x-ray_2024} indicates the bulk-like nature of the altermagnetic order.}
\end{figure*}

\section{\label{sec:Experimental}Measuring XMCD in bulk Manganese Telluride}
To probe the bulk altermagnetism in MnTe, we fabricate a lamella of thickness ranging from $150 - 200\,\rm{nm}$ extracted from bulk single crystal MnTe  with focused ion beam (FIB). The crystal {(shown in Figure \ref{fig:fig2}(a))} was grown with chemical vapour transport \cite{deMelo_91_MnTegrowth}, and the lamella was extracted such that it is oriented perpendicular to the $c$-axis, with the $a,b$ axes lying in the plane. {The relative orientation of the $a$ and $b$ axes with respect to the lamella are indicated in Figure \ref{fig:fig3}(b,c).}
More details of the crystal growth, characterisation, and lamella preparation are given in the Appendix. 
The three dimensional crystal structure around the Mn atoms, and the different sublattices, are shown in Figure \ref{fig:fig2}. }
{Due to the hexagonal crystal structure and anisotropy, the antiferromagnetic moments lie in the $ab$ plane, as shown in Figure \ref{fig:fig2}(b,c), with three easy axes along which the Néel vector in the antiferromagnetic domains typically orients \cite{Bogdanov1998_LTP, Kriegner2016_NatCommn}.} The two {antiferromagnetic} sublattices of MnTe are related by mirror symmetry, rather than a simple translation or inversion. As a result, the two domains  are related by time reversal symmetry, satisfying the symmetry classification of altermagnetism.

Quantitative measurements of the XMCD signal through the thickness of the lamella were obtained with scanning transmission X-ray microscopy (STXM) at the MAXYMUS beamline at BESSY II. 
The STXM setup is shown schematically in Figure \ref{fig:fig3}(a).
X-ray optics are used to focus the X-ray probe to a spot size of {25}\,nm. The sample is then scanned with respect to the beam, and the transmitted photons recorded at each point. 
By performing scans with circular left and right polarised light {in the vicinity of the Mn $L_{\textrm{2,3}}$ edges}, one can obtain a projection of the XMCD signal with nanoscale resolution{, that integrates through the thickness of the lamella parallel to the $c$-axis of MnTe.}  Measurements were performed at a temperature of 100 K, sufficiently below the N\'eel temperature of 307 K, where both XMCD and AHE signals have previously been measured \cite{hariki_x-ray_2024,amin_altermagnets_2024,Gonzales_AHE_MnTe}

The presence of XMCD in the lamella was confirmed by performing XMCD STXM with an X-ray energy of $640.30\,\rm{eV}$, {$0.75\,\rm{eV}$ below the energy corresponding to maximum absorption at the Mn $L_{\mathrm{3}}$ absorption edge. }
The electronic and XMCD images of the MnTe lamella are given in Figure \ref{fig:fig3}(b) and (c), respectively (see Appendix for details of XMCD calculation).
In the XMCD image, one can immediately see bright and dark XMCD contrast varying throughout the lamella.

\section{\label{sec:Results}XMCD spectroscopic analysis}

To determine whether our observed XMCD signal is associated with altermagnetic order, we perform XMCD spectro-microscopy across the Mn-$L_\mathrm{2,3}$ edges, following the energy evolution of the XMCD contrast of the domain features. A series of XMCD projections of a zoomed region of the lamella taken across the $L_{\mathrm{3}}$ and $L_{\mathrm{2}}$ edges are shown in Figure \ref{fig:fig4}(a) and (b), respectively. As one traverses both edges, one can see the domain features changing contrast a total of eight times across the $L_{\mathrm{3}}$ edge, and twice across the $L_{\mathrm{2}}$ edge, consistent with the oscillating contrast of altermagnetic order in MnTe \cite{hariki_x-ray_2024}. An XMCD spectrum, measured in a single domain region with an energy step size of $0.05\,\rm{eV}$ in the vicinity of the $L_{\mathrm{2,3}}$ edges is plotted {with a dashed line} in Figure \ref{fig:fig4}(c) alongside the absorption spectrum (solid line), and confirms this oscillating signal.  
We can extract XMCD signals from the set of images measured across the edge by comparing the signal within domains of opposite contrast. 
This XMCD contrast is calculated quantitatively as a percentage of the maximum absorption of the material at the $L_\mathrm{3}$ edge, defined as $A_{\mathrm{max}}$ in Figure \ref{fig:fig4}(c), and is plotted alongside the spectrum in Figure \ref{fig:fig4}(c).
Not only does the XMCD contrast qualitatively agree with spectra calculated with DFT, with the XMCD contrast exhibiting nine peaks of alternating sign across the $L_{\mathrm{3}}$ edge \cite{hariki_x-ray_2024}: a remarkable agreement is observed in the quantitative signal as well. 
Indeed, previous spectroscopic measurements reported an order of magnitude difference between the magnitude of the predicted, and measured altermagnetic XMCD signal {\cite{hariki_x-ray_2024}}. 
Here, we observe a maximum XMCD contrast of {$1.8\pm0.3\% $ of $A_{\mathrm{max}}$,  the maximum absorption, which agrees well with the predicted XMCD contrast of $1.8\% $ of $ A_{\mathrm{max}}$ \cite{hariki_x-ray_2024}.

{This quantitative agreement between the theoretical calculations of XMCD, and the spectra measured here allows us to draw two main conclusions.  
First, by measuring the spectrum with nanoscale spatial resolution, we remove the effect of averaging over domains, and instead can extract a signal that can be quantitatively compared with theory. 
Second, the quantitative agreement between our data and the DFT calculations, when compared to the absorption of the material, allows us to determine whether the XMCD signal originates from ordering at the surface of the sample or exists through the bulk of the lamella.
Indeed, if we consider the two cases of a surface altermagnetic order, and a bulk altermagnetic order, shown schematically in Figure \ref{fig:fig4}(d), a large discrepancy in the measured value of the XMCD would be expected. For a surface effect, where we assume the surface order to be limited to {the order of} $10\,\rm{nm}$ of the surface, the XMCD signal would be expected to less than { 0.2\% } of the total absorption of the sample. 
For the bulk order, and a continuous domain through the sample, the XMCD signal should correspond to approximately { 1-2\%  } of the total absorption, as we observe here. 
As a result, we can conclude that in our crystalline MnTe sample, the altermagnetic order exists in the bulk of the lamella of {$200\,\rm{nm}$ thickness}.

\section{Altermagnetic configuration}

Having confirmed that the bulk MnTe exhibits altermagnetic order, we next consider the domain configuration present in the lamella. As observed in Figure\,\ref{fig:fig3}(c), a number of regions with bright and dark XMCD can be observed. In MnTe, the XMCD is closely linked to the symmetry of the magnetic order. In particular, {the XMCD is expected to vary as a function of the orientation of the Néel vector in the $ab$ plane and exhibit three-fold rotational symmetry \cite{amin_altermagnets_2024}}, as shown schematically in the inset of Figure \ref{fig:fig5}(b). As a result, a change in XMCD contrast indicates a rotation of the Néel vector, and the formation of altermagnetic domains. 

\begin{figure*}
    \centering
    \includegraphics{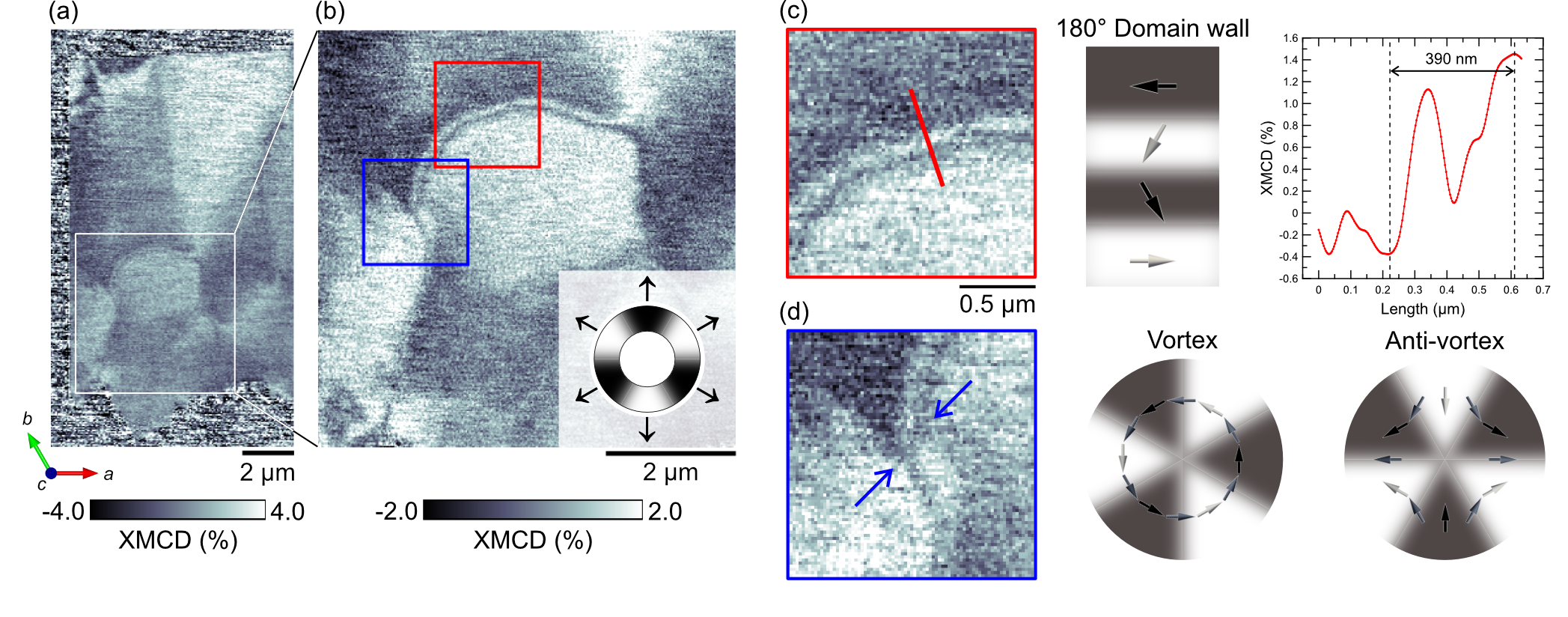}
     \caption{Altermagnetic domain configurations and nanotextures. (a) Overview of the domain configuration within the lamella. (b) A zoom of the region indicated by the white box in (a) reveals a number of nanoscale features within the configuration, where the XMCD changes rapidly. The XMCD contrast depends on the in-plane orientation of the Néel vector with three-fold symmetry, as indicated by the contrast wheel in the inset. (c) A closer look at the configuration highlighted by the red box in b) reveals an elongated structure indicated by the red line, across which the XMCD contrast reverses from white to black to white to black. Considering the angular dependence of the XMCD, this contrast is consistent with the presence of a 180$^\circ$ Néel domain wall, as shown schematically. This oscillating contrast can be seen in a line profile of the XMCD taken across the domain wall, revealing the domain wall width to be 390\,nm. (d) Considering the configuration highlighted by the blue box in (b), one can see two features in the XMCD (indicated by blue arrows) that resemble a cartwheel-like contrast, with three-fold rotational symmetry. Such contrast is consistent with the presence of a winding texture, which could be a vortex or an antivortex, as shown schematically. However, with the XMCD signal alone, it is not possible to determine exactly which configuration is present.  }\label{fig:fig5}
\end{figure*}

\subsection{Altermagnetic domains}
We first consider the overall  configuration of the lamella, which appears to host a rather complex domain structure.  {Notably, the domain size is generally on the order of micrometres in size. This is significantly larger than the domains observed in thin films grown by MBE \cite{amin_altermagnets_2024}, and could be due to the difference in microstructure or strain between the single crystal lamella measured here and the thin films. 
Indeed, in our lamella, we notice that the {domain size varies between the centre of the lamella – where the domains are relatively large, on the order of 5\,\textmu$\rm{ m}$ – and the edges, where smaller domains on the order of 1\,\textmu$\rm{ m}$ form.
}
To understand this variation in the size of the domains, we can consider two main differences between the edge and the center. First, the presence of surface effects such as slight differences in the strain distribution, that have been shown to influence antiferromagnetic \cite{Folven_nanostruct_AFM,Reimers_nanostruct_AFM2024} and altermagnetic configurations \cite{amin_altermagnets_2024}.
Secondly, due to the fabrication process, the edges of the lamella are more likely to contain defects. %at the edges of a FIB-patterned lamella, the additional exposure to the focused ion beam during the fabrication process can lead to an increase in the presence of defects, though this is typically limited to the top 10 nm of a surface. 
As in an antiferromagnet,  strain, disorder or structural defects often lead to the formation of domains that otherwise would not be energetically favourable \cite{Gomonay_strain_AFMdomains_2002}. We therefore conclude that both edge-strain effects, and microstructural defects could play a role in determining the size of the altermagnetic domains. }

In addition to the change in the size of the domains, we note that the XMCD is not a homogeneous bright or dark contrast throughout the sample - which would be associated with the Néel vector aligning along the three easy axes in the $ab$ plane. The presence of the intermediate grey contrast could be consistent with either the orientation of the Néel vector pointing away from a principal axis \cite{hariki_x-ray_2024}, consistent with a low anisotropy in the plane, and the smooth variation of the magnetic moment orientations. Alternatively, as we are integrating through the sample, the contrast could arise from an unresolved three dimensional domain structure \cite{donnelly2017vectorTOMO}.

\subsection{Altermagnetic textures}
As well as the presence of magnetic domains that can be identified as regions of homogeneous XMCD contrast, there exist a number of features in the XMCD projection that indicate the presence of nanoscale textures of the Néel vector. In the first work harnessing XMCD-photoelectron emission microscopy to image the magnetic configuration of an altermagnet, vortices, antivortices and domain walls of the Néel vector could be identified by their XMCD and XMLD contrast in thin film MnTe, and indeed could even be controlled by the micropatterning of the thin film \cite{amin_altermagnets_2024}.
Here, we consider a number of features in the region of the lamella indicated by the white box in Figure\,\ref{fig:fig5}(a), and shown in more detail in Figure\,\ref{fig:fig5}(b-d). Specifically, we consider two features in more detail: first, an elongated wall that appears between black and white domains, highlighted by the red box in Figure\,\ref{fig:fig5}(b) and shown in (c). Across the boundary between the two domains, the XMCD contrast does not switch simply from black to white, but rather alternates black - white - black - white across a distance of {390 nm}. When considering the dependence of the XMCD contrast on the direction of the Néel vector in the schematic, one can see that such contrast is consistent with a 180$^\circ$ domain wall of the Néel vector with a width of {390 nm}, across which the XMCD contrast switches due to the reversal of the sublattice moments.  Multiple such 180$^\circ$ domain walls were observed in this configuration, as well as other configurations within additional lamellae, indicating that this is a stable domain texture for this material.

In addition to the presence of domain walls, we also observed a number of more complex textures in the XMCD contrast, as highlighted by the blue box in Figure\,\ref{fig:fig5}(b) and indicated by arrows in (d). In particular, isolated configurations with a three-fold symmetry cartwheel-like contrast are observed, with the XMCD changing contrast six times as one follows a line around the central points (indicated by blue arrows). Again, considering the symmetry of the XMCD contrast, such features are consistent with the presence of topological textures that exhibit 360$^\circ$ winding of the Néel vector, such as the vortices and antivortices that are shown schematically in Figure\,\ref{fig:fig5}(d). Although XMCD measurements alone are insufficient to distinguish between vortices and antivortices of opposite winding, the observed contrast indicates the presence of such topological magnetic textures in bulk altermagnets. The presence of such winding textures has been observed in hexagonal compensated magnets in thin films - both MnTe \cite{amin_altermagnets_2024}, and hematite \cite{jani2024_hematitemerons}. Their observation here highlights that such textures can also spontaneously occur in lamellae extracted from single crystals, indicating that they are at least a metastable configuration.

\section{\label{sec:Conclusion}Conclusion}
In conclusion, we have obtained experimental evidence for altermagnetic order in bulk MnTe. By performing nanoscale XMCD imaging in transmission on a {$150-200\,\rm{nm}$} thick lamella extracted from a bulk crystal, we observe altermagnetic domains{, domain walls and winding topological textures } with a spectroscopic signature characteristic of altermagnetic order in bulk MnTe. Comparison of the XMCD signal with DFT calculations \cite{hariki_x-ray_2024} reveals an excellent quantitative agreement, indicating that this altermagnetic order exists through the thickness of the lamella, and is not confined to the surface, confirming the bulk nature of the state. 

This experimental evidence of altermagnetic ordering in the bulk provides an important insight for future studies of altermagnetic phenomena. The consistency of the XMCD spectrum measured here from bulk crystals with thin films grown by MBE indicates that the order is robust to strain effects originating from the substrate in such thin films, providing evidence for the intrinsic nature of altermagnetism in MnTe. With these results, we further demonstrate that XMCD spectroscopy is a robust, quantitative technique to probe altermagnetic order, and when combined with nanoscale imaging, provides a route to probe individual altermagnetic domains within complex configurations. 
This ability to investigate, and characterise altermagnetic order in bulk crystals will represent an important tool for the exploration, and understanding, of altermagnetism across a wide range of candidate materials, of key importance for the development of future technologies.

Finally, the measurement of XMCD in transmission allows for the probing of bulk order through the thickness of a sample. Going forward, this could be combined with vector dichroic tomographic methods \cite{donnelly2017vectorTOMO,Apseros2024XLDtomo} to map 3D altermagnetic configurations. We note that here, a lamella of $200\,\rm{nm}$ thickness was measured, which is associated with 20\% transmission on resonance, thus limiting the on resonance imaging to thin systems. However, due to the oscillating nature of the XMCD signal, considerable XMCD is also found in the pre-edge as seen in Figure \ref{fig:fig4}(c), where there is significantly higher transmission, offering a route to the mapping of 3D altermagnetic configurations in micrometre-thick systems \cite{Jeffrey2024}. Lastly, we note that one of the advantages of transmission X-ray imaging is that due to its photon-in, photon-out nature, it can be easily combined with \textit{in situ} measurements. The combination with transport measurements, electrical stimuli or the application of magnetic fields would provide a route to determine the microscopic origin of the anomalous Hall effect measured in altermagnets, as well as explore the manipulation of altermagnetic textures on the nanoscale. 

\appendix
\section{Growth of MnTe single crystals {and X-ray characterisation}}
First, MnTe was synthesised by a direct reaction of the elements in an equimolar ratio Mn (pieces, Chempur 99.99\%, powdered directly before use) and Te (powder, Alfa Aesar 99.999\%) with the addition of {1.5 mg/ml ampoule volume of} iodine (Alfa Aesar 99.998\%) at 500 °C in evacuated quartz glass tubes for 10 days.
Following this, MnTe crystals were grown from the resulting microcrystalline powder by chemical vapour transport in a temperature gradient from 700 °C (source) to 650 °C (sink) with iodine (Alfa Aesar 99.998 \%) 1.5 mg/ml as a transport agent.

{The crystal structure of the MnTe was identified by powder X-ray diffraction. For the first investigation, a sample with an irregular shape and dimensions of approximately 0.05 mm was mechanically extracted from a grown crystal. For characterisation using the powder method, several crystals were pulverised and the resulting powder was placed in a glass capillary with a diameter of 0.3\,mm. Data was collected at the ID-22 beamline of European Synchrotron Radiation Facility. The hexagonal structure of NiAs type was confirmed with the following parameters: space group $P6_{3}/mmc$ No, 194, $a = 4.1483(1)\,\rm{\AA}$, $c = 6.7162(3) \rm{\AA}$, Mn at $2a$ (0, 0, 0), Te at $2c$ ($\frac{1}{3}$, $\frac{2}{3}$, $\frac{1}{4}$). The refinement of the occupation of both positions (Mn and Te) did not indicate any deviation from the stoichiometric 1:1 composition.}

\section{Preparation of lamella with focused ion beam}
The MnTe lamella was extracted from a bulk single crystal using a focused Ga ion beam. A standard lift-out method, utilizing an \textit{in situ} micromanipulator, was employed, and the sample was mounted onto a copper transmission electron microscopy grid, where it was further thinned to a {$150-200\,\rm{nm}$}. The lamella was oriented with the $c$ axis normal to the plane.
The orientation of the lamella with respect to the crystallographic orientation was determined by performing Laue measurements on the crystal from which the lamella was extracted. 

\section{Scanning transmission X-ray microscopy}
Scanning transmission X-ray microscopy was carried out at the MAXYMUS endstation at the BESSY-II synchrotron operated by Helmholtz-Zentrum Berlin für Materialien und Energie. Using a Fresnel zone plate and order selecting aperture, the X-ray beam was focused to a spot size of $\approx$ 25 nm. The sample was cooled in the absence of external magnetic fields to 100 K using an \textit{in situ} helium cryostat. To acquire images, the sample was scanned through the X-ray beam using a piezoelectric motor stage, with the transmission measured pixel by pixel using an avalanche photodiode. Sequential images of both positive and negative X-ray circular polarization were acquired as a function of energy across the Mn $L_{2,3}$ edges, in order to track and exploit the X-ray magnetic circular dichroism.

\section{Calculation of XMCD}
The X-ray magnetic circular dichroism images presented in this paper were calculated in the following way.
The transmitted intensity of circularly polarised X-rays ($I^\pm$), measured in the experiment, is given by 
\begin{equation}
I^\pm = I_0 e^{-\mu^\pm d},
\end{equation}
where $I_0$ is the incident intensity, $\mu^\pm$ is the energy and magnetism-dependent attenuation coefficient for positive (negative) circular polarized X-rays, and $d$ is the thickness of the sample. The normalized intensity is defined as 
\begin{equation}
I^\pm_{\textrm{norm}} = I^\pm/I_0, 
\end{equation}
The XMCD signal is calculated as:  
\begin{equation}
\text{XMCD} = 100 \times \frac{(I^+_{\textrm{norm}} - I^-_{\textrm{norm}})}{A_{\textrm{max}}} \%
\end{equation}

Here, $A_{\textrm{max}}$ is the maximum absorption of the material at the $L_{3}$ edge (See Figure \ref{fig:fig4}). Calculating the XMCD as a percentage of this value allows us to obtain thickness-independent values for the XMCD. These values can then be quantitatively compared with different energies, and furthermore can be compared with predicted XMCD values from DFT calculations \cite{hariki_x-ray_2024}.

\begin{figure*}
\includegraphics[width=0.9\textwidth]{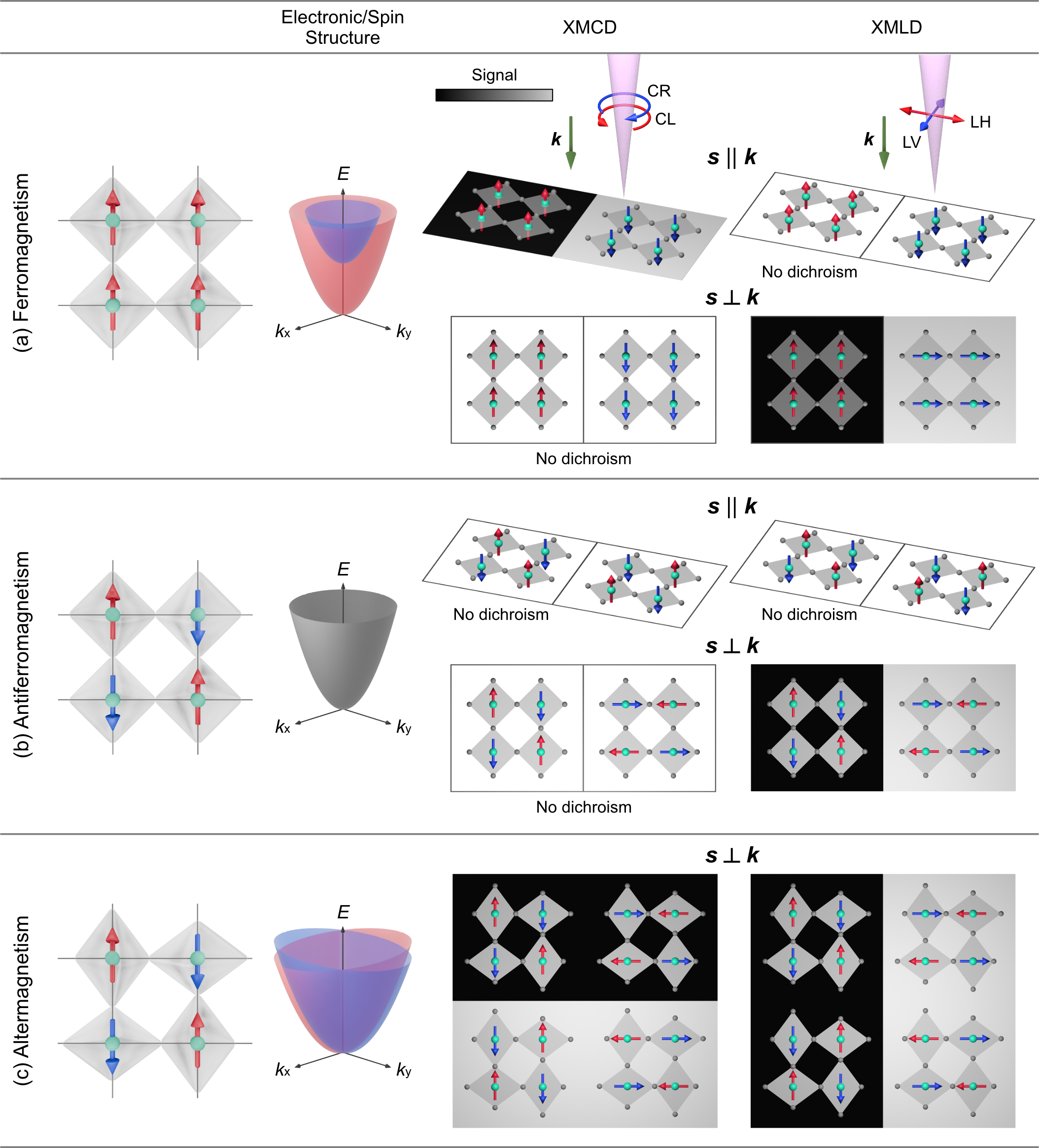}%
\caption{\label{fig:fig6} { Mapping magnetic domains {in ferromagnets, antiferromagnets and altermagnets} using polarized X-rays.
The spin configuration and local crystal environments  of ordered magnetic ions (left), the electronic and spin structure in momentum space (centre) and the expected XMCD and XMLD contrast for different domain configurations (right) is shown for ferromagnetism (a), antiferromagnetism (b) and  $d$-wave altermagnetism (c). Throughout, red and blue indicate opposite spin orientations, while the black, and white, and grey planes in the XMCD/XMLD schematics indicate positive, negative and zero XMCD or XMLD intensity, respectively. 
(a) In ferromagnets, neighbouring spins have parallel alignment, associated with a well-defined splitting in momentum space. For the spins orientated collinear to the X-rays ($\bm{s} \parallel \bm{k}$), there is a non-zero X-ray magnetic circular dichroism (XMCD), but no X-ray magnetic linear dichroism (XMLD). For the spins lying in-plane  ($\bm{s} \perp \bm{k}$), there is no XMCD, however XMLD distinguishes between perpendicular spin configurations.
(b) For collinear antiferromagnets with spins aligned to the X-rays ($\bm{s} \parallel \bm{k}$), there is no XMCD or XMLD. For in-plane spin configurations ($\bm{s} \perp \bm{k}$), there is no XMCD signal associated with this state, however XMLD distinguishes between perpendicular spin configurations. 
(c) In d-wave altermagnets shown here, at least four types of magnetic domains form due to the combination of antiferromagnetic spin configurations, and the inequivalent local environments. For the case of X-rays propagating perpendicular to the plane in which the spins lie, there is both non-zero XMCD - that probes the relative orientation of the spins with respect to the local environment, and linear dichroism, which probes the relative orientation of the spins with respect to the linear polarisation of the light. }
}
\end{figure*}
% If you have acknowledgments, this puts in the proper section head.
\begin{acknowledgments}
We thank the Helmholtz Zentrum Berlin fur Materialien und Energie {and the ESRF} for the allocation of synchrotron radiation beamtime.  R.Y., L.T., M.D.P.M., J.C.C.F. and C.D. acknowledge funding from the Max Planck Society Lise Meitner Excellence Program and funding from the European Research Council (ERC) under the ERC Starting Grant No. 3DNANOQUANT 101116043. L.T. and M.D.P.M. acknowledge the support of the Alexander von Humboldt Foundation. S.F. acknowledges funding from the Swiss National Science Foundation under grant no. IZSEZ0{\_}223146. funding from the Swiss National Science Foundation under grant no. IZSEZ0{\_}223146
The authors thank Seunghyun Khim for support with Lau\'e measurements and Vicky Hasse for support with the synthesis of the samples. 
\end{acknowledgments}

% Create the reference section using BibTeX:

% \bibliography{ref}
%

\end{document}